%% file: repthm-QPL-FINREV-10-1-17.tex
\documentclass[submission,copyright,creativecommons]{eptcs}
\providecommand{\event}{QPL 2017} 
\usepackage{breakurl}             
\usepackage{underscore}           

\title{A Shortcut from Categorical Quantum Theory to Convex Operational Theories}
\author{Alexander Wilce
\institute{Department of Mathematics, Susquehanna University}
}
\def\titlerunning{A Shortcut from Categorical Quantum Theory to Convex Operational Theories}
\def\authorrunning{A. Wilce}

\usepackage{amsfonts, amssymb, amsmath} 
\input xy
\xyoption{all}
\usepackage{tikz,amsmath}

\input{discardingtree}

\newcommand{\tempout}[1]{{}}

\newcounter{thaler}
\newenvironment{mlist}{\begin{list}{\arabic{thaler}}%
{\usecounter{thaler}
\setlength{\rightmargin}{\leftmargin}
\topsep=0pt
\itemsep=0pt
\parskip=0pt
\parsep=0pt
}}{\end{list}}

\newtheorem{theorem}{Theorem}[section]
\newtheorem{lemma}[theorem]{Lemma}
\newtheorem{proposition}[theorem]{Proposition}
\newtheorem{corollary}[theorem]{Corollary}
\newtheorem{example}[theorem]{Example}

\newtheorem{definition}[theorem]{Definition}

\newcommand{\C}{{\boldsymbol {\cal C}}}

\newcommand{\K}{{\boldsymbol {\cal K}}}

\newcommand{\R}{{\mathbb R}}

\newcommand{\id}{\text{id}}
\newcommand{\op}{\text{op}}
\renewcommand{\H}{{\boldsymbol {\mathcal H}}}
\newcommand{\Tr}{\text{Tr}}
\newcommand{\tr}{\text{Tr}}

\newcommand{\ostar}{\circledast}

\newcommand{\FdHilb}{{\bf FdHilb}}

\newcommand{\FRel}{{\bf FRel}}
\newcommand{\Rel}{{\bf Rel}}
\newcommand{\Set}{{\bf Set}}
\newcommand{\Rvec}{{\bf RVec}}
\newcommand{\OrdLin}{{\bf OrdLin}}
\newcommand{\Hilb}{{\bf Hilb}}

\newcommand{\Pinj}{{\bf Pinj}}

\newcommand{\B}{{\cal B}}

\renewcommand{\L}{{\cal L}}
\newcommand{\sa}{\mbox{sa}}

\newcommand{\e}{\mbox{{\bf ev}}}
\newcommand{\ev}{\e}

\renewcommand{\hat}{\widehat}
\renewcommand{\tilde}{\widetilde}
\newcommand{\V}{V_{o}}
\newcommand{\Vin}{V_{\infty}}
\newcommand{\E}{V^{\#}_{o}}
\newcommand{\Eu}{V_{\infty}^{\#}}
\newcommand{\wt}{\widetilde}
\newcommand{\wh}[1]{[ #1 ]}

\renewcommand{\Tr}{\text{Tr}}

\newcommand{\OrDP}{{\bf OrDP}}

\begin{document}

\maketitle

\begin{abstract} 
This paper charts a very direct path between the categorical approach to quantum mechanics, due to Abramsky and Coecke, and the older convex-operational approach based on ordered vector spaces (recently reincarnated as ``generalized probabilistic theories"). In the former, the objects of a symmetric monoidal category $\mathcal C$ are understood to represent physical systems and morphisms, physical processes. Elements of the monoid ${\mathcal C}(I,I)$ are interpreted somewhat metaphorically as probabilities. Any monoid homomorphism from the scalars of a symmetric monoidal category $\C$  gives rise to a  covariant functor $V_o$ from $\C$ to a category of dual-pairs of ordered vector spaces. 
Specifying a natural transformation $u : V_o \rightarrow 1$ (where $1$ is the trivial such functor) allows 
us to identify normalized states, and, thus, to regard the image category $V_o(\C)$ as consisting of concrete 
operational models. In this case, if $A$ and $B$ are objects in $\C$, then $V_o(A \otimes B)$ defines a non-signaling composite of $V_o(A)$ and $V_o(B)$. Provided either that $\C$ satisfies a ``local tomography" condition, or that $\C$ is compact closed, this defines a symmetric monoidal structure 
on the image category, and makes $V_o$ a (strict) monoidal functor. 
\end{abstract}

\section{Introduction} 
 

\noindent {\it Note: This is a revised 
and expanded version of notes privately circulated around 2010.  Proposition 4.3 on representations of compact closed 
categories is new.}

\bigskip

\noindent In the categorical quantum mechanics of Abramsky and Coecke \cite{ACHbk}, physical theories are understood as symmetric monoidal categories, with physical systems as objects, physical processes as morphisms, and the monoidal structure allowing for the  composition of systems and processes ``in parallel". The scalars in such a category play  the role, in a somewhat metaphorical sense, of probabilities. An older tradition, going back at least to the work of Ludwig \cite{Ludwig}, Davies and Lewis \cite{Davies-Lewis} and Edwards \cite{Edwards}, models a physical system more concretely in terms of a dual pair of ordered vector spaces, one 
spanned by the system's states, the other 
by ``effects" (essentially, measurement outcomes), with the duality prescribing the probability with which any given effect will occur in any given state. These concrete ``operational" models can be combined by means of various possible non-signaling products \cite{BW}. 

Several attempts have been made to connect the two approaches. On one hand, several authors \cite{Coecke-Selby, Gogioso-Scandolo, Jacobs, Tull} have considered categories equipped with additional structure mirroring some of the structure found in the more concrete approach. 
On another hand, symmetric monoidal categories of concrete operational models have been constructed and studied in, e.g., \cite{BDW, Wilce}.   
This paper attempts to link the categorical 
and operational approaches in a much more direct way, by constructing {\em representations} of (essentially arbitrary) symmetric monoidal categories {\em as} monoidal categories of concrete probabilistic models.
The basic idea is simply to posit a homomorphism 
from the commutative monoid  of scalars of the category, to the multiplicative monoid of non-negative real numbers, 
providing an interpretation of (some) scalars as genuine probabilities.

Depending on the model category $\C$ one has in mind, a morphism $\alpha : I \rightarrow A$ from the tensor unit $I$ to an object $A \in \C$ may be taken to represent a pure state, a mixed state, or possibly a sub-normalized 
--- or even, totally un-normalized --- state of the system $A$. 
So far as possible, one would like to be able to deal with all of these cases in a reasonably uniform manner.  To this end, I first construct, for a given monoid homomorphism $p : \C(I,I) \rightarrow \R_+$, a more or less obvious functor $V_o : \C \rightarrow \OrdLin$ from $\C$ into  the category of ordered real vector spaces and positive linear mappings. There is a well-defined product $V_o(A), V_o(B) \mapsto V_o(A \otimes B)$ on the image category $V_o(\C)$, satisfying 
certain desiderata for a composite of convex operational models (Proposition 4.1). 
Under an additional local tomography assumption (satisfied by all of the usual examples, but which one would certainly like to weaken), {\em or} if $\C$ is compact closed, this product makes $V_o(\C)$ monoidal, and $V_o$, a strict monoidal functor 
(Propositions 4.2 and 4.3).

This much depends only on the monoid homomorphism $p$. To distinguish between normalized and non-normalized states,  a bit more is required, namely, for each object $A \in \C$, a posited {\em unit effect} $u_A \in V_o(A)^{\ast}$. 
This is meant to represents the trivial event that is certain to occur. Accordingly, one {\em defines} a normalized state to be an element of $\alpha \in V_o(A)_+$ with $u_A(\alpha) = 1$. 
Such states form a convex set $\Omega_{o}(A)$, which is a base for the positive cone $V_o(A)_+$ (that is, every element of the cone is a multiple of normalized state). 

I will actually require a little more still. First, in order for normalization to behave correctly under the composition, 
I will ask that 
$u_{A \otimes B} = u_{A} \otimes u_{B}$ 
for all $A, B \in \C$. In order to allow us to interpret elements of $\C(A,I)$ as being at least multiples 
of effects, I also require that, 
for every $a \in \C(A,I)$, the corresponding evaluation functional on $V_{o}(A)$ 
be dominated by 
a positive multiple of $u$, i.e, 
\begin{equation} 
\forall a \in \C(A,I) \ \exists t \geq 0 \ a \leq t u_A. 
\end{equation} 
 Let $V_{o}^{\#}(A)$ denote the (by construction, separating) ordered subspace of $V_{o}(A)^{\ast}$ spanned by 
the evaluation functionals associated with elements of $\C(A,I)$, 
The unit effect $u_A$ will often belong to $V_{o}^{\#}$; when it does, (\theequation) makes it an 
order unit for $V_{o}^{\#}(A)$, and the triple $(V_o(A),V_{o}^{\#}(A), u_A)$ is then a convex operational 
model as defined, e.g., in \cite{BDW}.  

In infinite-dimensional settings, where in general $u_A$ will not belong to $\E(A)$, 
one can also consider a number of ways of ``completing" $\V(A)$. Three of these, 
denoted $V_1(A)$, $Vin(A)$ and $M(A)$, are considered in Section 5. Since the choice of order-unit $u$ is not canonical, these are not functorial on $\C$. 
However, the choice of $u$ determines a sub-category, $\C_u$, of $\C$, having the same objects, but with a restricted set of ``physical" morphisms that respect the order unit. 
$V_1$, $\Vin$ and $M$ are the object parts of functors $\C_u \rightarrow \OrdLin$, and I {\em conjecture} that 
$(V(A \otimes B),V^{\#}(A \otimes B), u_{AB})$ is a  
``non-signaling" composite of $(V(A),V^{\#}(A), u_A)$ and $(V(B),V^{\#}(B), u_B)$. 

These constructions are illustrated for a number of model categories $\C$, including the categories of finite-dimensional Hilbert spaces (where we recover the expected thing) and the category of finite sets and binary relations.  Although I 
have tried to leave ample room for infinite-dimensional examples,  I've avoided the head-on engagement with the linear-topological issues that this project will ultimately require.

\section{Preliminaries}  

I denote the category of (all!) real vector spaces and linear mappings by $\Rvec$; however, for $V, W \in \Rvec$, 
I write 
$\L(V,W)$, rather than $\Rvec(V,W)$, for the space of linear mappings $V \rightarrow W$.  
I write $V^{\ast}$ for the full (algebraic) dual of $V$, 
$V \otimes W$ for the algebraic tensor product of $V$ and $W$, and ${\cal B}(V,W)$ for the space of bilinear forms $V \times W \rightarrow {\Bbb R}$.   
By an {\em ordered vector space}, I mean a real vector space $V$ equipped with a convex, pointed, generating cone $V_{+}$. Any space of the form ${\Bbb R}^{X}$, $X$ a set, will be 
understood to be ordered pointwise on $X$. 
If $V, W$ are ordered linear spaces, a mapping $f : V \rightarrow W$ is {\em positive} iff $f(V_+) \subseteq W_+$. 
The {\em dual cone} of an ordered linear space $V$ is the cone $V^{\ast}_{+}$ consisting of all positive linear functionals on $V$ (where $\R$ is understood to have 
its usual order).  The span 
of $V^{\ast}_{+}$ in $V^{\ast}$ is called the {\em order dual} of $V$, and denoted $V^{\star}$ 
I write $\OrdLin$ for the category of ordered linear spaces and positive linear mappings.

\noindent{\bf Representations.} A (real, linear) {\em representation} of a category $\C$ is simply a covariant functor 
$V : \C \rightarrow \Rvec$. 
There are two standard functors $\Set \rightarrow \Rvec$, one contravariant, given on objects by $X \mapsto {\Bbb R}^{X}$, and the other covariant, given on objects by $X \mapsto {\Bbb R}^{[X]}$, where the latter is the vector space 
generated by $X$, or, equivalently, the space of finitely non-zero functions on $X$.  
Thus, given a reference object $I \in \C$, we have basic representations 
$\C \rightarrow \Rvec$ and $\C^{\op} \rightarrow \Rvec$ given on objects by $A \mapsto {\Bbb R}^{\C(A,I)}$ and by $A \mapsto {\Bbb R}^{[\C(I,A)]}$, respectively. 
A representation $V$ is {\em finite-dimensional} iff $V(A)$ is finite-dimensional for 
every object $A \in \C$. By a representation of a {\em symmetric monoidal} category $\C$, I mean a functor 
$V : \C \rightarrow \Rvec$ that is symmetric monoidal with respect to {\em some} well-defined 
monoidal product on the image category. Of course, given the functor, there is only one candidate for this product. The following is obvious, but worth stating explicitly. 

\begin{lemma} Let $V : \C \rightarrow \Rvec$ be a functor such that the operations (i) $V(A), V(B) \rightarrow V(A \otimes B)$ and (ii) $V(\phi), V(\psi) \rightarrow V(\phi \otimes \psi)$ 
are well-defined\footnote{That is to say: if 
$V(A) = V(A')$ and $V(B) = V(B')$, then $V(A \otimes B) = V(A' \otimes B')$, and similarly for morphisms.} for $A,B \in \C$ and $\phi \in \C(A,C)$, $\psi \in \C(B,D)$.  Then the image category $V(\C)$ is monoidal with respect to the product given by $V(A) \ostar V(B) := V(A \otimes B)$ (with 
associators, left and right units, and swap morphisms carried over from $\C$). With respect to this structure, 
$V$ is a strict monoidal functor.\footnote{Henceforth, ``monoidal" will always mean ``strict monoidal".} \end{lemma}

Notice that this imposes no special linear or multilinear structure on the product in $V(\C)$. For instance, we would like  
to have, at a minimum, canonical bilinear maps $V(A) \times V(B) \rightarrow V(A \otimes B)$.  We would also probably 
want to require that $V(I) = \R$. I return to these points below. \\

\noindent{\bf Dual pairs and Convex Operational Models.} For present purposes, we may define a {\em dual pair of 
ordered vector spaces} --- an {\em ordered dual pair}, for short --- as a pair $(V,V^{\#})$, where $V$ is an 
ordered vector space and $V^{\#}$ is a subspace of $V^{\ast}$, ordered by a cone $V^{\#}_{+}$ contained 
in the dual cone $V^{\ast}_+$. In other words, if $b \in V^{\ast}_+$ and $\alpha \in V_+$, 
$b(\alpha) \geq 0$ (but it may be that $b(\alpha) \geq 0$ for all $\alpha \in V_+$, yet $b \not \in V^{\#}_+$). 
I will also assume, without further comment, that $V^{\#}$ is separating, i.e., that if $\alpha \in V$ and $b(\alpha) = 0$ for all $b \in V^{\#}$, then $\alpha = 0$. 
The following language is borrowed from \cite{BDW}, but the idea is essentially the same one proposed by Ludwig \cite{Ludwig}, Davies and Lewis \cite{Davies-Lewis}, Edwards \cite{Edwards} and others in the 1960s and 1970s as a general framework for post-classical probabilistic physics. 

\begin{definition} {\em 
A {\em convex operational model} (COM) is a triple $(V,V^{\#},u)$ where 
$(V,V^{\#})$ is an ordered dual pair and  $u \in V^{\#}$ is a chosen order unit. [NOT DEF'D??]\footnote{The so-called 
{\em no-restriction hypothesis} \cite{NoRes}, usually stated informally as the requirement that all ``mathematically  possible" effects be physically realizable, amounts to the requirement that $V^{\#} = V^{\star}$, the order-dual of $V$. 
One wants to avoid this very strong assumption wherever possible.} } 
\end{definition}

A COM gives us a very general environment in which to discuss probabilistic concepts. 
An element $\alpha$ of $V_+$ with $u(\alpha) = 1$ is a {\em normalized state} of the model. An {\em effect} of the model is an element $a$ of $V^{\#}_+$ with $a \leq u$;
equivalently, $a(\alpha) \leq 1$ for all normalized states $\alpha$.  Effects represent (mathematically) possible measurement outcomes: if $a$ is an effect and $\alpha$ is a normalized state, $a(\alpha)$ is interpreted as the {\em probability} that $a$ will occur (if measured) in state $\alpha$.

\begin{example}[Motivating Examples] {\em (a) Let $(S,\Sigma)$ be a measure space. Let $V = M(S,\Sigma)$, the space of all countably-additive measures on $S$, and $V^{\#} = B(S,\Sigma)$, the space of all bounded measurable functions on $S$, 
with the duality given by $f(\mu) := \int_{S} f(s) d\mu(s)$ for every $f \in B(S,\Sigma)$ and $\mu \in M(S,\Sigma)$. 
 The constant function $1$ serves as the order unit. 
(b) Let $\H$ be any Hilbert space: take $V = \L_{1}(\H)$, the space of trace-class self-adjoint operators on $\H$, and $V^{\#} =  \L_{\sa}(\H)$, the space of all bounded Hermitian operators on $\H$,
with the duality given by the trace, i.e, if $\alpha$ is a trace-class operator and $a$, any bounded operator, $\sigma(\alpha,a) = \Tr(a\alpha)$. The identity operator on $\H$ serves as the order unit. 
}
\end{example}

\noindent{\bf Composite Systems}  
Suppose that $(V,V^{\#},v)$ and $(W,W^{\#},w)$ are two (convex operational) models, representing two physical systems. In attempting to form a reasonable model of a composite system , the most obvious construction 
--- $(V \otimes W, V^{\#} \otimes W^{\#}, v \otimes w)$ --- is rarely appropriate. Certainly in infinite dimensions, one will typically need to pass from $V \otimes W$ to some appropriate linear-topological completion; but even 
where $V$ and $W$ are finite-dimensional, there are at least two further issues:
\begin{itemize} 
\item[$\bullet$] There is no one canonical choice for the cones $(V \otimes W)_+$ and $(V^{\#} \otimes W^{\#})_{+}$: there do exist minimal and maximal 
tensor cones \cite{BW}, but in the quantum-mechanical examples, these yield the wrong things.
\item[$\bullet$] As is well known, in the case of real or quaternionic quantum models, where $V = \L_{\sa}(\H)$ and $W = \L_{\sa}(\K)$ for finite-dimensional real or quaternionic Hilbert spaces $\H$ and $\K$, 
one finds, upon counting dimensions, that $\L_{\sa}(\H \otimes \K) \not \simeq \L_{\sa}(\H) \otimes \L_{\sa}(\K)$. 
\end{itemize} 
How, then, {\em should} one define a composite of two models? At a minimum, one wants to be able to construct {\em joint measurements} and prepare the systems independently in any two states. One also 
wants to be able to form, from a joint state $\omega \in VW$, the {\em conditional} state of, say, $W$, given the result of a measurement (an effect) on the first system.  
This suggests the following definitions \cite{Wilce}. (A bilinear mapping is {\em positive} iff it takes positive values on positive arguments.)

\begin{definition} A {\em composite} of two ordered dual pairs $(V,V^{\#})$ and $(W,W^{\#})$ is an ordered dual pair $(VW,(VW)^{\#})$, together with positive 
blinear mappings $\ostar : V \times W \rightarrow VW$ and $\pi : V^{\#} \times W^{\#} \rightarrow (VW)^{\#}$ 
such that
\begin{itemize} 
\item[(a)] $\pi(a,b)(\alpha \ostar \beta) = a(\alpha) b(\beta)$ 
for all $\alpha \in V, \beta \in W, a \in V^{\#}$ and $b \in W^{\#}$.   
\item[(b)] $\pi(-, b)(\omega) \in V_+$ and $\pi(a, -)(\omega) \in W_+$ for 
all $a \in V^{\#}_{+}$ and $b \in W^{\#}_{+}$
\end{itemize}
A composite of COMs $(V,V^{\#},v)$ and $(W,W^{\#},w)$ is a COM $(VW, (VW)^{\#}, vw)$ where 
$(VW, (VW)^{\#})$ is a composite of ordered dual pairs, and $vw = v \otimes w$.
\end{definition}

If $a$ and $b$ are effects in $V^{\#}$ and $W^{\#}$, respectively, then $\pi(a,b) =: a \otimes b$ is an effect in $(VW)^{\#}$, called a {\em product effect}, 
and interpreted as the result of measuring $a$ and $b$ jointly on the systems represented by $(V,V^{\#},u)$ and $(W,W^{\#},v)$. 

Restricting the dual mapping $\pi^{\ast} : (VW)^{\# \ast} \rightarrow \B(V^{\#}, W^{\#})$ 
to (the canonical image of) $VW$ in $(VW)^{\# \ast}$, we have a canonical bilinear mapping 
\[\Lambda : V \times W \rightarrow \B(V^{\#}, W^{\#})\]
We then have $\Lambda(\omega)(a,b) = (a \otimes b)(\omega)$. Accordingly, I refer to $\Lambda$ as the {\em localization mapping}. The idea is that if $\omega \in VW$ is a state of the composite system, then
$\Lambda(\omega)$ is object assigning joint probabilities to pairs of outcomes of ``local" measurements associated with the component systems, represented by $V$ and $W$, respectively.  

\begin{definition}{\em 
A composite $(VW,(VW)^{\#}, vw)$ {\em locally tomographic} 
iff $\Lambda$ is injective. }
\end{definition} 

In other words, $(VW, (VW)^{\#}, vw)$ is locally tomographic iff local joint probabilities suffice to determine the 
joint stat of $V$ and $W$. \\

\noindent{\em Remark:} The bilinearity of $\pi$ (or of $\Lambda$) is equivalent to the ``no-signaling" condition. If $E = \{a_i\}$ is an observable of the COM $(V,V^{\#},v)$, i.e, a set of effects summing to the unit $v$, 
and $\omega$ is a state of $\Omega(VW)$, then  the {\em marginal state} of $B$, given this observable, is 
defined, $\forall b \in W^{\#}$, by 
$\omega_{E}(b)  \ = \ \sum_{i} \Lambda(\omega)(a_i \otimes b) = \Lambda(\omega)(\sum_i a_i \otimes b) = \Lambda(\omega)(v \otimes b)$,
which is evidently independent of $E$. The interpretation is that the probability of observing an effect $b$ on the system corresponding to $(W,W^{\#},w)$ is 
independent of which measurement we make on the system corresponding to $(V,V^{\#},v)$. This works equally well in the other direction. We thus 
have well-defined marginal states $\omega_1 = \Lambda(\omega)(v, ~\cdot~ )$ and $\omega_2 = \Lambda(\omega)( ~\cdot~ , w)$. Condition (b) in Definition 2.4 guarantees that these actually belong to $W_+$ and to $V_+$, respectively, and not just to $(W^{\#})^{\ast}_{+}$ and $(V^{\#})^{\ast}_{+}$. 
\\

Given a  functor $V : \C \rightarrow \OrdLin$ in which $V(I) = {\Bbb R}$, we have $V(a) \in V(A)^{\ast}$ for all $a \in \C(A,I)$. Letting $V^{\#}(A)$ denote the span, in $V(A)^{\ast}$, of the functionals 
$V(a)$, $a \in \C(A,I)$, we have a functor $A \mapsto (V(A),V^{\#}(A))$ from $\C$ to the category $\OrDP$ of real dual pairs. We should like this to be monoidal, in the sense that the obvious (and only) candidate for a monoidal product on the image category be well-defined, but also, yield products of ordered dual pairs, in the sense defined above, {\em and} interact with the monoidal structure carried over from that of $\C$ in a sensible way. 

The following definition attempts to make these requirements precise. 

\begin{definition} {\em A {\em  monoidal ordered linear representation} of a symmetric monoidal category $\C$, is a functor 
$V : \C \rightarrow \OrdLin$, such that (i) the constructions 
\[V(A), V(B) \mapsto V(A \otimes B) \ \ \mbox{and} \ \ V(\phi), V(\psi) \mapsto V(\phi \otimes \psi) : V(A \otimes C) \rightarrow W(B \otimes D)\]
with $\phi \in \C(A,B), \psi \in \C(C,D)$, are well-defined,
together with(ii) for all objects $A$ and $B$, bilinear mappings $\ostar_{A,B} : V(A) \times V(B) \rightarrow V(A \otimes B)$ and $\pi_{A,B} : V^{\#}(A) \times V^{\#}(B) \rightarrow \V^{\#}(A \otimes B)$ 
making $(V(A \otimes B), V^{\#}(A \otimes B))$  a composite in the sense of Definition 2.4, of $(V(A),V^{\#}(A))$ and $(V(B),V^{\#}(B))$, and such that (iii) 
\[V(\alpha) \ostar V(\beta) = V(\alpha \otimes \beta) \ \text{and} \ 
\pi(V(a), V(b)) = V(a \otimes b)\]
for all $\alpha \in \C(I,A)$, $a \in \C(A,I)$, $\beta \in \C(I,B)$, and $b \in \C(B,I)$. 
}
 \end{definition}

\tempout{
\noindent{\em Remarks} (a) By Lemma 2.1, the image category $V(\C)$ of $\C$ under such a representation will be a symmetric monoidal category, and $V$, a (strict) monoidal functor.  Suppose now that $V(I) = \R$. Then every morphism 
$\alpha \in \C(I,A)$ gives rise to a vector $V(\alpha)(1) \in V(A)$. If $V(A)$ is spanned by vectors of this form, for all $A \in \C$, then 
 condition (iii) is equivalent to the requirement that, for all $A, B, C, D \in \C$ and all morphisms $\phi : A \rightarrow C$ and $\psi : B \rightarrow D$, 
$V(\phi \otimes \psi)(v \ostar w) = V(\phi)(v) \ostar V(\psi)(w)$
for all vectors $v \in V(A)$ and $w \in V(B)$. 

(b) Definition 2.6 could be stated more abstractly. Both of the mappings $(V \ostar V) : A, B \mapsto V(A \ostar B) := V(A \otimes B)$ and $(V \otimes V) : A, B \mapsto V(A) \otimes V(B)$ are (object parts of) functors from $\C \times \C$ into $\Rvec$. If we equip $V(A) \otimes V(B)$ with the 
minimal tensor cone, consisting of linear combinations $\sum_{i} t_i v_i \otimes w_i$ where $v_i \in V(A)_+$, $w_i \in V(B)_+$, and with coefficients $t_i \geq 0$, then a positive bilinear mapping 
$ V(A) \times V(B) \rightarrow V(A \otimes B)$ extends uniquely to a positive linear mapping $V(A) \otimes V(B) \rightarrow V(A \otimes B)$. In this way, we can regard both 
$V \ostar V$ and $V \otimes V$ as functors $\C \times \C \rightarrow \OrdLin$. From this point of view, the family of positive bilinear mappings $\ostar_{A,B}$ posited in part (ii) are the components of 
a natural transformation $V \otimes V \stackrel{\cdot}{\rightarrow} V \ostar V$. Similarly, the we can regard the mappings $\pi_{A,B}$ as components of a natural transformation 
from $V^{\#} \times V^{\#}$ to $V^{\#} \circ \otimes$, where, again, $V^{\#}(A) \otimes V^{\#}(B)$ has the minimal tensor cone. 
}

\tempout{\noindent{\em Remark:} This could be stated still more precisely. I have in mind that $\Lambda$ and $\ostar$ should be part of the data. Thus, we're talking about a triple $(V,\ostar,\Lambda)$ where 
$V : \C \rightarrow \Rvec$, and $\ostar$ and $\Lambda$ are natural transormations  
$\Lambda_A : W(A \otimes - ) \rightarrow \B(W^{\#}(A),W^{\#}( - ))$ and $\ostar_{A} : W(A) \otimes W(-) \rightarrow W(A \otimes - )$. (I imagine there's a standard notion of bifunctor, and of natural transformation between bifunctors, which could be applied here.) 
}

\section{The Representation $V_o$}

There is a particularly simple, and canonical, representation of any category $\C$ in $\OrdLin$. 
As discussed above, there is a ``largest" contravariant linearization functor $\Set \rightarrow \Rvec$, namely, 
$X \mapsto {\Bbb R}^{X}$, $f \mapsto f^{\ast}$, 
where, if $f : X \rightarrow Y$, $f^{\ast} : {\Bbb R}^{Y} \rightarrow {\Bbb R}^{X}$ is the linear  mapping taking $\beta \in {\Bbb R}^{Y}$ to 
$f^{\ast}(\beta) = \beta \circ f$. Composing this with the contravariant 
$\Set$-valued functor $A \mapsto \C(A,I)$, $\phi \mapsto \phi^{\ast}$, 
where, again, $\phi^{\ast}$ is defined by $\phi^{\ast} (a) = a \circ \phi$ for all $a \in \C(A,I)$, gives us a covariant functor 
$\C \rightarrow \Rvec$, taking each object $A$ to the (huge) vector space ${\Bbb R}^{\C(A,I)}$ and each morphism $\phi \in \C(A,B)$ to the 
linear mapping $\phi_{\ast} : {\Bbb R}^{\C(A,I)} \rightarrow {\Bbb R}^{\C(B,I)}$ given by 
\[\phi_{\ast}(\alpha)(b) = \alpha(\phi^{\ast}(b)) = \alpha(b \circ \phi)\]
for all $\alpha \in {\Bbb R}^{\C(A,I)}$ and all $b \in \C(B,I)$. (Of course, we can do the same using any vector space, or, for that matter, any set, in place of ${\Bbb R}$.)
With respect to the natural pointwise ordering on spaces of the form $\R^{X}$, 
the linear mappings defined above are positive. Thus, we can regard the functor just defined as taking $\C$ to $\OrdLin$, where the latter is the category of ordered linear spaces and positive linear mappings. 

Suppose now that $\C$ is a symmetric monoidal category (SMC) with tensor unit $I$.
Let $S = \C(I,I)$ be the monoid of scalars in $\C$, and let $p : S \rightarrow {\Bbb R}_{+}$ be a monoid homomorphism (where 
we regard ${\Bbb R}_{+}$ as a monoid under multiplication). For each $\alpha \in \C(I,A)$, let $\wh{\alpha} \in {\Bbb R}^{\C(A,I)}$ 
be the function defined by 
\[\wh{\alpha}(a) = p(a \circ \alpha)\]
for all $a \in \C(A,I)$. Ultimately, we wish to be able to identify those $\alpha \in \C(I,A)$ and those 
$a \in \C(A,I)$ that correspond to actual physical states and effects (or events), and, for such a pair, 
to regard $\wh{\alpha}(a)$ as the {\em probability} that 
the effect $a$ occurs when the system $A$ is in state $\alpha$. This will require some further attention to 
questions of normalization, which we'll return to in section 5. Meanwhile, we are now in a position to 
represent elements of $\C(I,A)$ as elements of the positive cone of a ordered vector space: 

\begin{definition}
 Let $V_o(A)$ denote the linear span of the vectors $\wh{\alpha} \in \R^{\C(A,I)}$, ordered pointwise.\footnote{This notation suppresses the dependence of $V_o$ on the choice of monoid homomorphism $p$. Should it 
become necessary to track this dependence, we can write $V^{p}_{o}$ or something of the sort.} 
\end{definition} 

If $\phi \in \C(A,B)$, we have 
\[\phi_{\ast}(\wh{\alpha})(b) = \wh{\alpha}(b \circ \phi) = p(b \circ \phi \circ \alpha) = \wh{\phi \circ \alpha} (b)\]
for all $b \in \C(B,I)$. Thus, we may regard $\phi_{\ast}$ as mapping $V_o(A)$ to $V_o(B)$. Writing $V_o(\phi)$ for $\phi_{\ast}$, thus 
restricted and co-restricted, we have a functor $V_o : \C \rightarrow \OrdLin$. \\

\noindent{\em Remark:} The functor $V_o$ will sometimes be degenerate. For instance, if $\C$ is a meet semi-lattice, with $a \otimes b = a \wedge b$ and $I = 1$ (the top element of $\C$), we have 
$\C(I,I) = \{1\}$, and there is a unique monoid homomorphism $p : S \rightarrow {\Bbb R}_{+}$. However, for $a \not = I$, $\C(I,a) = \emptyset$, whence, $V_o(a)$ is again empty.

\begin{lemma} For any SMC $\C$, $V_o(I)$ is canonically isomorphic to ${\Bbb R}$. \end{lemma}

\noindent {\em Proof:} For all $s, t \in \C(I,I)$, we have $\wh{s}(t) = p(s \circ t) = p(s) p(t)$. Thus, $\wh{s} = p(s)p \in {\Bbb R}^{S} = V(S)$, whence, 
$V_o(I)$ is the one-dimensional span of $p \in V_{o}(I)$.  $\Box$\\ 

Up to this canonical isomorphism $V_o(I) \simeq {\Bbb R}$, we can now identify $V_o(s)$ with $p(s)$ for all $s \in \C(I,I)$. 
Also, for $\alpha \in \C(I,A)$, we have $V_o(\alpha) : V_o(I) \rightarrow V_o(A)$. Up to the canonical isomorphism 
$V_o(I) \simeq {\Bbb R}$, we have $V_o(\alpha)(1) = [\alpha]$. Similarly, if $a \in \C(A,I)$, have $V_o(a) : 
V_o(A) \rightarrow V_o(I) \simeq {\Bbb R}$, i.e., $V_o(a) \in V_o(A)^{\ast}$, given by $V_o(a)(\rho)(1) = \rho(a)$. 
Denoting the evaluation functional $\rho \mapsto \rho(a)$ by $\hat{a}$, we have $V_o(a) = \hat{a}$ (again, up to the identification of $V_o(I)$ with ${\Bbb R}$). Note that $\hat{a}([\alpha]) = [\alpha](a) = V_o(a \circ \alpha)$. \\
Note also that the evaluation functional $\hat{a} \in V_o(A)^{\ast}$ is positive for all $a \in \C(A,I)$. 
 
\begin{definition} Let $\E(A)$ denote the span of these functionals $[a]$, ordered by the cone they generate. 
\end{definition}

For every $\phi \in \C(A,B)$, we have a dual mapping $(V_o \phi)^{\#} : \E(B) \rightarrow \E(A)$, and hence, dual to this, a mapping $V_o(\phi) := (\phi^{\#})^{\ast} : V_o(A) \rightarrow V_o(B)$. Thus, the ordered dual pair $(V_o(A), \E(A))$ is functorial in $A$. 

In what follows I will live a bit dangerously and simply write $a$ for $\hat{a} \in \E(A)$, leaving it to context to disambiguate usage. (In particular, 
we are {\em not} assuming that $a \mapsto \hat{a}$ is injective.)

\begin{example}[Finite-dimensional Hilbert spaces]{\em Let $\FdHilb$ be the category of finite-dimensional complex Hilbert spaces and linear mappings. We have have $\C(I,A) \simeq \C(A,I) \simeq A$, where $\alpha : I \rightarrow A$ is determined by the vector $\alpha(1) \in A$, while $a \in \C(A,I)$ is determined by $a(x) = \langle v_a, x \rangle$ for a unique $v_a \in A$. The monoid $\C(A,A)$ is (isomorphic to) $({\Bbb C}, \cdot)$. Let $p : {\Bbb C} \rightarrow {\Bbb R}_{+}$ be given by $p(z) = |z|^2$. If $\alpha \in \C(A,I)$ and $a \in \C(A,I)$ then $\alpha(a) = p(a \circ \alpha) = |\langle v_a, \alpha(1) \rangle |^2$, which gives the usual quantum-mechanical transition probability. From this 
point forward, we identify $a \in \C(A,I)$ with $v_a \in A$; then we may interpret $\wh{\alpha} \in {\Bbb R}^{A}$, via $\wh{\alpha}(a) = |\langle a | \alpha(1) \rangle|^2 
 = \langle \alpha(1) \odot \alpha(1) a, a \rangle$ --- in other words, $\wh{\alpha}$ is the quadratic form associated with the rank-one operator 
 $\alpha(1) \odot \alpha(1)$.\footnote{Here, $a \odot b : \H \rightarrow \H$ is given by $(a \odot b)x = \langle x,b \rangle a$.}  It 
 follows that $V_o(A)$ is the space of (quadratic forms associated with) Hermitian operators on $A$. We also have $V_{o}^{\#}$ the span of rank-one Hermitian operators --- in our finite-dimensional setting, then, $V_{o}^{\#}(A) \simeq V_o(A)$. 
 
If $\phi \in \C(A,B) = \L(A,B)$, we have $v_{b \circ \phi} = v_{\phi^{\ast}(b)}$, i.e., $b \circ \phi = \phi^{\ast}(b)$. Hence, 
\[V_{o}(\phi)(\wh{\alpha})(b) = \wh{\alpha}(b \circ \phi) = \langle (\alpha(1) \odot \alpha(1)) \phi^{\ast}(b), \phi^{\ast}(b) \rangle = \langle \phi (\alpha(1) \odot \alpha(1)) \phi^{\ast} b, b \rangle,\]
so that $V_{o}(\phi)(\wh{\alpha}) = \wh{\phi(\alpha)}$; note that this also shows that $V_o(\phi) : \rho \mapsto \phi \rho \phi^{\ast}$ for all $\rho \in V_o(A) \simeq \L_{h}(A)$, i.e., $V$ implements the 
usual lifting of linear mappings from $A$ to $B$ to linear mappings from $\L_h(A)$ to $\L_h(B)$. 
 }\end{example}

\begin{example}[Arbitrary Hilbert spaces]
{\em Let $\Hilb$ the category of all separable complex Hilbert spaces, and bounded linear mappings. Again, we have $\C(A,I) \simeq A$. Here $V_o(A)$ is the space of finite-rank Hermitian 
operators on $A$. 
The same computation as above shows that, for a bounded linear mapping $\phi : A \rightarrow B$, $V_o(\phi) : V_o(A) \rightarrow V_o(B)$ is the conjugation mapping $\rho \mapsto \phi \rho \phi^{\ast}$. 
}\end{example}

\begin{example}[Relations]
{\em Let $\C = \Rel$, the category of sets and relations. The tensor unit is the one-point set $I = \{\ast\}$, so that $S = {\cal P}(I \times I) = \{\emptyset, \{(\ast,\ast)\}\} \simeq {\cal P}(I)$. 
Let's identify this with $\{0,1\} \subseteq {\Bbb R}_{+}$. We also have, for every $A \in \C$, isomorphisms $\C(A,I) \simeq \C(I,A) \simeq {\cal P}(A)$, with $\alpha \in \C(I,A)$ corresponding to $\alpha(\ast) \subseteq A$ and $a \in \C(A,I)$, to $a^{-1}(\ast) \subseteq A$. Let $p : S = \{0,1\} \rightarrow {\Bbb R}$ be the obvious injection. Then for all $a, \alpha \in {\cal P}(A)$, regarded as elements of $\C(A,I)$ and $\C(I,A)$, respectively, we have 
\[p(a \circ \alpha) = \left \{ \begin{array}{cl} 1 & \mbox{if} \ a \cap \alpha \not = \emptyset \\ 0 & \text{otherwise} \end{array} \right .\]
Thus, $\wh{\alpha} \in {\Bbb R}^{{\cal P}(A)}$ is the characteristic function of the set $[\alpha] = \{ a \subseteq A | a \cap \alpha \not = \emptyset\}$. We can regard this as a kind of {\em possibility measure} on ${\cal P}(A)$, in the sense that $\wh{\alpha}(a) = 1$ iff 
$a$ is possible, given that $\alpha$ is certain. $V_o(A)$ is the span of these possibility measures in ${\Bbb R}^{{\cal P}(A)}$ --- a space it 
would be nice to characterize more directly.} \end{example} 

\begin{example}[Categories with Very Small Hom Sets]{\em Let $\C$ be any SMC such that $\C(A,B)$ is finite for all objects $A, B \in \C$ --- for instance, any sub-category of the category of finite sets and 
relations. Let $S = \C(I,I)$, and let $R : S \rightarrow {\Bbb R}^{S}$ be the usual right action, given by $R_s(f)(x) = f(xs)$ for all $x,s \in S$ and all $f \in {\Bbb R}^{S}$. Since $R^{S}$ is finite-dimensional, 
the mapping $R_s$ is linear, and $R_{s_1 s_2} = R_{s_1} R_{s_2}$, we have a canonical 
monoid homomorphism $p : S \rightarrow {\Bbb R}_{+}$, namely $p(s) = |\det R_{s}|$, and hence, a canonical representation $V_o$.} 
\end{example}

\tempout{
\begin{example}[Partial Injections]{\em Let $C = \Pinj$, the category of sets and partial injective mappings. This is monoidal with $I$ the one-element set $\{\ast\}$ and $A \otimes B = A \times B$. We have $\C(A,I) \simeq A \simeq \C(I,A)$, with $\alpha \in \C(I,A)$ corresponding to $\alpha(0) \in A$ and $a \in \C(A,I)$ coresponding to $\alpha^{-1}(0)$. The scalar mpnoid is $\{0,1\}$. Letting 
$p : \{0,1\} \rightarrow {\Bbb R}_{+}$ be the inclusion mapping, we have $[\alpha](a) = 1$ if $\alpha = a$, and $0$ otherwise --- which is to say, $[\alpha]$ is the point-mass $\delta_{\alpha(0)} \in {\Bbb R}^{A}$. Hence, $V_o(A) = {\Bbb R}^{[A]}$, the space of finitely supported real-valued functions on $A$. $V_{o}^{\#}(A)$ is also isomorphic to ${\Bbb R}^{[A]}$.  }\end{example}

\noindent{\em Question:} If $W : \C \rightarrow \Rvec$ is any representation of $\C$ with $\dim(W(I)) = 1$, then we obtain another functor $\C \rightarrow \Rvec$ by composing $W$ with the canonical covariant functor 
$\Phi : \Rvec \rightarrow \Rvec$ given by $M \mapsto \L(W(I),M)$, $T  \in \L(M,N) \mapsto ( - \circ T) : \L(W(I),M) \rightarrow \L(W(I),N)$. We have a natural isomorphism $\L(W(I),W(I)) \simeq {\Bbb R}$, so this functor satisfies 
the conditions above. {\em When is this again a monoidal functor?} 
}

\section{Monoidality} 

We now adddress the questions of how $V_o$ interacts with the monoidal structure of $\C$.  We begin with the 
observation that the construction 
$V_o(A), V_o(B) \ \mapsto \ V_o(A \otimes B)$ 
is well-defined: the assignment $A \mapsto \C(A,I)$ is injective, and we can read off 
$\C(A,I)$ as the domain of any function in $V(A) = {\Bbb R}^{\C(A,I)}$.  Thus, given two spaces $V_o(A)$ and $V_o(B)$ in $V_o(\C)$, we can unambigously define  $V_o(A) \ostar V_o(B) := V_o(A \otimes B)$.  
Ultimately, we wish to invoke Lemma 2.1 to conclude that $V_o$ is the object part of a monoidal structure 
on $V_o(\C)$, with respect to which $V_o$ a monoidal functor. This requires that the construction 
\[V_o(\phi), V_o(\psi) \ \mapsto \ V_o(\phi \otimes \psi)\]
also be well-defined for all morphisms $\phi, \psi$ in $\C$. 
To insure this, additional constraints on $\C$ seem to be needed. One sufficient condition is that (i) the tensor unit $I$ be {\em separating}, meaning that the functor $\C( - , I)$
is injective on morphisms, and (ii) the monoid homomorphism $p : S \rightarrow {\Bbb R}$ is also injective. 
However, the injectivity of $p$ is a very strong constraint, not satisfied, for example, in the case of $\FdHilb$. Thus, we need to say something more in order to secure the monoidality of $V_o$.  

Before turning to this, however, we show that $V_o(A \otimes B)$ is indeed a composite of $V_o(A)$ and $V_o(B)$ --- or, more exactly, that the dual pair $(V_o(A \otimes B), \E(A \otimes B))$ is a composite of the dual pairs 
$(V_o(A),\E(A))$ and $(V_o(B),\E(B))$ ---  in the sense of Definition 2.2.
Recall \cite{BDW} that in any symmetric monoidal category, we have, for every $\omega \in \C(I, A \otimes B)$, 
a natural mapping $\wt{\omega} : \C(B,I) \rightarrow \C(I,A)$, given by  $\tilde{\omega}(b) = (\id_{A} \otimes b) \circ \omega$. 
Dually, if $f  \in \C(A \otimes B, I)$, there is 
a natural mapping $\wt{f} : \C(I,A) \rightarrow \C(B,I)$ given by $\tilde{f}(a) = f \circ (a \otimes \id_{B})$. 

\begin{proposition} 
For any objects $A$ and $B$ of $\C$, there exist canonical positive bilinear mappings  
\[\ostar : V_o(A) \times V_o(B) \rightarrow V_o(A \otimes B) 
\ \ \mbox{and} \ \ \pi : \E(A) \times \E(B) \rightarrow \E(A \otimes B)\]
making $(V_o(A \otimes B), V^{\#}_{o}(A \otimes B))$ a composite of ordered dual pairs in the sense of 
Definition 2.4. 
\end{proposition} 

\noindent{\em Proof:} There is only one candidate for $\ostar$: it must be the bilinear extension --- unique if it exists! --- of the mapping 
\[[\alpha], [\beta] \mapsto [\alpha] \ostar [\beta] := [\alpha \otimes \beta].\]
To see that this last is well-defined, let $\alpha \in \C(I,A)$, $\beta \in \C(I,B)$ and $f \in \C(A \otimes B,I)$: 
then 
\begin{equation} 
[\alpha \otimes \beta](f)  
 =  p(f \circ (\alpha \otimes \beta)) 
 =  p((f \circ (\id_A \otimes \beta)) \circ \alpha) 
 =  p(\wt{f}(\beta) \circ \alpha) 
 =  [\alpha](\wt{f}(\beta)).
\end{equation}
This depends only on $[\alpha]$. In a similar way, one sees that $[\alpha \otimes \beta] =  [\beta](\wt{f}(\alpha))$. Thus 
$[\alpha \otimes \beta]$ depends only on $[\alpha]$ for fixed $\beta$, and only on $[\beta]$ for fixed $\alpha$ --- and hence, only on $[\alpha]$ and $[\beta]$. 

Next, we must show that $[\alpha], [\beta] \mapsto [\alpha \otimes \beta]$ extends to a bilinear mapping $V_o(A) \times V_o(B) \rightarrow V_o(A \otimes B)$ (which 
will automatically be positive, if it exists).  This 
amounts to showing that $\sum_i t_i [\alpha_i], \sum_j s_j [\beta_j] \mapsto \sum_{i,j} s_i t_j [\alpha_i \otimes \beta_j]$ is well-defined. 
Let $\rho = \sum_i t_i [\alpha_i \otimes \beta] \in V_o(A \otimes B)$. Then, 
for every $f \in \C(A \otimes B,I)$, using (3), we have 
\[\rho(f) = \sum_{i} t_i [\alpha_i \otimes \beta](f) = \sum_i t_i [\alpha_i](\wt{f}(\beta))\]
which depends only on $\sum_i t_i [\alpha_i] \in V_o(A)$. A similar argument shows that $\sum_i s_j [\alpha \otimes \beta_j]$ depends only on $\sum_j s_j [\beta_j]$ for fixed $\alpha$, whence, $\sum_{i,j} t_i s_j [\alpha_i \otimes \beta_j]$ depends only on $\sum_i t_i [\alpha_i]$ and $\sum_j s_j [\beta_j]$. 

So much for $\ostar$. We next need to show that the mapping $\otimes : \C(A,I) \times \C(B,I) \rightarrow \C(A \otimes B, I)$ extends uniquely to a positive bilinear mapping 
\[\pi : \E(A) \times \E(B) \rightarrow \E(A \otimes B)\]
with $\pi(a,b) (\alpha \ostar \beta) = a(\alpha) b (\beta)$ 
For all $\mu \in \V(A \otimes B) \leq \R^{\C(A \otimes B, I)}$, define 
\[\tilde{\mu} : \C(A,I) \rightarrow \R^{\C(B,I)}\]
by $\tilde{\mu}(a)(b) = \mu(a \otimes b)$.  Note that the mapping 
$\mu \mapsto \tilde{\mu}$ is linear. {\em Claim:} $\tilde{\mu}(a) \in \V(B)_+$. It's enough 
to check this for $\mu$ having the form $[\omega]$, $\omega \in \C(I, A \otimes B)$, as 
these span $\V(A \otimes B)$.  
But now 
\[\tilde{\mu}(a)(b) = p((a \otimes b) \circ \omega) = p(b \circ (a \otimes \id_{B}) \circ \omega).\]
Letting $\hat{\omega}(a) := (a \otimes \id_{B}) \circ \omega \in \C(I, B)$, we have 
\[\tilde{\mu}(a) = [\hat{\omega}(a)] \in \V(B)_+\]
as claimed.  
 We now have a mapping $\tilde{\mu} : \C(A,I) \rightarrow \V(B)$. Dualizing, 
we have a linear mapping $\tilde{\mu}^{\ast} : \V(B)^{\ast} \rightarrow \R^{\C(A,I)}$. 
If  $a \in \C(A,I) \subseteq \V(B)^{\ast}$, we have $\tilde{\mu}^{\ast}(b)(a) = \mu(a \otimes b)$. 
Arguing in the same way as above, we see that $\tilde{\mu}^{\ast}(b) \in \V(A)$.  Dualizing again, 
we have a positive linear mapping $\V(A)^{\ast} \rightarrow \V(B)^{\ast \ast}$.  For 
$a \in \E(A) \leq \V(A)^{\ast}$, $\tilde{\mu}(a) \in \V(B)$, so by restriction we have a positive 
linear mapping $\tilde{\mu} : \E(A) \rightarrow \V(B)$. This gives us a bilinear mapping 
$\E(A) \times \E(B) \rightarrow \R$, which 
we shall also write as $\tilde{\mu}$, defined by $\tilde{\mu}(a,b) = \tilde{\mu}(a)(b)$. Now, as 
$\mu \mapsto \tilde{\mu}$ is linear, we have a positive linear mapping 
\[\V(A \otimes B) \rightarrow \B(\E(A),\E(B))\]
Dualizing for a final time, we have a bilinear mapping $\pi : \E(A) \times \E(B) \rightarrow \V(A \otimes B)^{\ast}$ 
given by 
\begin{equation} 
\pi(a, b) (\mu) = \tilde{\mu}(a,b) = \tilde{\mu}(a)(b) = \tilde{\mu}^{\ast}(b)(a).\end{equation}
If $a \in \C(A,I)$ and $b \in \C(B,I)$, we have $\pi(a,b)(\mu) = (a \otimes b)(\mu)$, so 
the range of $\pi$ lies in $\E(A \otimes B)$.  

It remains to verify conditions (ii)(a) and (ii)(b) of Definition 2.4. For the latter, notice that, by 
(\theequation), $\pi(a, - )(\mu) = \tilde{\mu}(a) \in V_o(B)_+$ and $\pi(a, - )(\mu) = \tilde{\mu}^{\ast}(b) \in V_o(A)_+$.   
For the former, observe that, 
for all $\alpha \in \C(I,A), a \in \C(A,I)$, $\beta \in \C(I,B)$ and $b \in \c(B,I)$,
\begin{eqnarray*}
\pi(a,b)([\alpha] \ostar [\beta])
& = & [\alpha \otimes \beta](a \otimes b)\\
& = & p((\alpha \otimes \beta) \circ (a \otimes b))\\
& = & p((\alpha \circ a) \otimes (\beta \circ b)) \\
& = & p(\alpha \circ a)p(\beta \circ b) = [\alpha](a)[\beta](b).  \ \ \Box \end{eqnarray*}

\tempout{
It remains to show that $\Lambda(\omega)(a, - ) \in V(B)_+$ and $\Lambda(\omega)(-, b) \in V(A)_+$ 
for all $a \in V^{\#}(A)_+$ and $b \in V^{\#}(B)_+$. A priori, $\Lambda(\omega)(a, - ) \in V^{\#}(B)^{\ast}$. 
But if $\omega \in \C(I,A \otimes B)$, $a \in \C(A,I)$ and $b \in \C(B,I)$, we have 
\[\Lambda(\omega)(a, -)(b) = \Lambda(\omega)(a,b) = p((a \otimes b) \circ \omega) 
= p(b \circ (a \otimes \id_B) \circ \omega) = [(a \otimes \id_B) \circ \omega)](b).\]
Hence, $\Lambda(\omega)(a, - ) = [(a \otimes \id_B) \circ \omega] \in V(B)_+$. Since 
vectors of the form $a \in \C(A,I)$ generate the cone $\E(A)_{+}$, it follows that 
$\Lambda(\omega)(a, - ) \in V_o(B)_+$ for all $a \in \E(A)_+$. Finally, since 
elements of the form $[\omega]$, $\omega \in \C(I,A \otimes B)$, span $V_o(A \otimes B)$, 
for all  $\omega \in V_o(A \otimes B)$ we have $\Lambda(\omega)(a,-) \in V_o(B)$. 
If $\omega \in V_o(A \otimes B)_+$, then $omega(a \otimes b) \geq 0$ for all 
$a \in \C(A,I)$ and $b \in \C(B,I)$, so $\Lambda(\omega)(a, -)(b) \geq 0$ 
for all $b \in \C(B,I)$, so $\Lambda(\omega) \in V_o(B)_+$. $\Box$\\}

We now return to the question of whether the product $V_o(A), V_o(B) \mapsto V_o(A) \ostar V_o(B)$ is the object part of a monoidal structure on $\C$. At present, I can't show that this is always the case. 
We do, however, have two sufficient conditions. One of these is local tomography of $V_o(\C)$:

\begin{proposition}  If the composite $(V_o(A \otimes B), \E(A \otimes B))$ is locally tomographic for all $A, B \in \C$, then $V_o$ is the object part of a monoidal 
 representation in the sense of Definition 2.3.  \end{proposition} 

\noindent{\em Proof sketch:}  Appealing to Lemma 2.1, we need only show that $V_o(\phi), V_o(\psi) \mapsto V_o(\phi \otimes \psi)$ is well-defined for $\phi \in \C(A,B)$ and $\psi \in \C(C,D)$ 
for all objects $A,B,C,D \in \C$.
 Let $\phi \in \C(A,B)$. We have a well-defined dual mapping $V_o(\phi)^{\#} : \E(B) \rightarrow \E(A)$, given by 
\[(V\phi)^\# (b)(\alpha) = V(\phi)(\alpha)(b) = p(b \circ \phi \circ \alpha).\]
Now, if $\C$ is locally tomographic, then $\Lambda : V(A \otimes B) \rightarrow B(\E(A),\E(B))$ is injective, so that 
$V(\phi \otimes C) (\omega)$ is determined by values of 
\begin{eqnarray*} V_o(\phi \otimes C)(\omega)(a \otimes b) & = & p((a \otimes b) \circ (\phi \otimes C) \circ \omega)\\
&  = & p(((a \circ \phi) \otimes b) \circ \omega)\\
& =  & V_o(\phi \otimes C)^{\#}(a \otimes b)(\omega) = (V_o(\phi)^{\#}(a) \otimes b)(\omega).\end{eqnarray*}
As the right-hand side depends only on $V_o(\phi)^{\#}$, and hence, on $V_{o}(\phi)$, rather than on $\phi$, the mapping $V_o(\phi) \mapsto V_o(\phi \otimes C) : V_o(A \otimes C) \rightarrow V_o(B \otimes C)$ is well-defined. A similar argument
shows that $V_o(A \otimes \psi)$ depends only on $V_o(\psi)$. Thus, we have a well-defined 
mapping $V_o(\phi), V_o(\psi) \mapsto V_o(\phi \otimes \psi)$. 
$\Box$ \\ 

As noted above, all of our benchmark categories -- $\FdHilb$, $\FRel$, etc. --- are locally tomographic. However, one can have $V_o$ monoidal in the absence of local tomography, as in the case of real Hilbert space. 
On the other hand, all of our {\em finite-dimensional} examples, including $\Rel$, are compact closed. Since 
all morphisms in such a category are represented by states, this is also sufficient:

\begin{proposition} If $\C$ is compact closed, then $V_o$ is the object part of a monoidal representation 
in the sense of Definition 2.3. \end{proposition} 

\noindent{\em Proof sketch:} Since $\C$ is compact closed, there is a natural mapping 
$\C(I,A^{\ast} \otimes B)$ to $\C(A,B)$ given by 
\[\omega \mapsto \ \hat{\omega} = (\epsilon_{A} \otimes \id_{B}) \circ (\id_{A} \otimes \omega).\]
Moreover, every morphism in $\C(A,B)$ arises in this fashion. 
If $\omega \in \C(I,A^{\ast} \otimes B)$ and $\mu \in \C(I,C^{\ast} \otimes D)$, then 
one finds that $\hat{\omega} \otimes \hat{\mu} 
= \hat{\omega \odot \mu}$ where 
$\omega \odot \mu := \tau \circ (\omega \otimes \mu)$  \ \ \mbox{and} \ \ 
and 
$\tau = \id_{A^{\ast}} \otimes \sigma_{B,C^{\ast}} \otimes \id_{D}$.
Now if $\omega, \omega' : I \rightarrow A^{\ast} \otimes B$ and $\mu : I \rightarrow C^{\ast} \otimes D$, 
$[\omega] = [\omega'] \Rightarrow [\omega \otimes \mu] = [\omega' \otimes \mu]$ (see Equation (1) in the proof of Proposition 4.1). 
Letting $\phi = (\epsilon_{(A \otimes B)^{\ast}} \otimes \epsilon_{C \otimes D}) \circ \tau$, 
we have 
\begin{eqnarray*}
V_o(\hat{\omega} \otimes \hat{\mu}) = V_o(\hat{\omega \odot \mu}) 
& = & V_o(\phi \circ (\omega \otimes \mu))\\
& = & V_o(\phi) \circ V_o(\omega \otimes \mu)\\
& = & V_o(\phi) \circ V_o(\omega' \otimes \mu) \\
& = & V_o (\phi \circ (\omega' \otimes \mu)) = V_o(\hat{\omega' \odot \mu}) = V_o(\hat{\omega'} \otimes \hat{\mu}).
\end{eqnarray*}
A similar computation in the other argument shows that, for 
$\phi \in \C(A,B)$ and $\psi \in \C(C,D)$, $V_o(\phi \otimes \psi)$ depends only 
on $V_o(\phi)$ and $V_o(\psi)$.  $\Box$

\section{Normalization} 

To this point, we have made no attempt to distinguish between normalized and non-normalized states. From the 
convex-operational point of view, only normalized and sub-normalized states represent actual states of affairs; 
super-normalized states are a mathematical convenience. In order to make this distinction in the present context, 
we introduce some new structure, namely, a choice, for each object $A \in \C$, of a positive functional 
$u_A$ that picks out the normalized states. This should be well-behaved in the following sense:

\begin{definition} 
{\em Let $V$ be a monoidal ordered representation of a symmetric monoidal category $\C$. 
An {\em unit} for $V$ is a choice, 
for each $A \in \C$, of a {\em strictly} positive functional $u_A \in V(A)^{\ast}$, such that 
\begin{mlist} 
\item[(i)] For every $a \in \C(A,I)$,  there exists some $t \in \R$, $t \geq 0$, such that $a \leq t u$;
\item[(ii)] for all $\alpha \in \C(I,A)$ and $\beta \in \C(I,B)$, 
$\displaystyle u_{A \otimes B}(\alpha \ostar \beta) = u_{A}(\alpha) u_{B}(\beta)$. 
\end{mlist} 
}
\end{definition}

\noindent{\em Remarks:} (a) If $V = V_o$, then the requirement that $u_A$ be strictly positive is 
redundant, as this follows from condition (i). (b) If $V(I) = \R$, we can interpret an unit as a natural transformation $u : V \rightarrow 1$, where $1$ is the trivial representation $1(A) = \R$ for all objects $A$ and 
$1(\phi) = \id_{\R}$ for all $\phi \in \C(A,B)$. (c)  $\C$ is a {\em discard category} \cite{Coecke-Selby} if every object $A \in \C$ is equipped with a morphism $\discard{A} : A \rightarrow I$ such that 
$\discard{A} \otimes \discard{B} = \discard{A \otimes B}$. In this case,  $u_A := V(\discard{A})$ will supply a unit, {\em provided} this satisfies condition (i).  \\

While the existence of a unit is not guaranteed, the usual examples have {\em canonical}  units. In $\FRel$, where $\C(A,{\ast}) \simeq \C({\ast},A) \simeq {\cal P}(A)$, there is a natural 
unit, namely $u_A = A$ --- or rather, 
$u_A(\wh{\alpha}) = \wh{\alpha}(A) = p(\alpha \cap A) = 1$ for all non-empty $\alpha \in {\cal P}(A)$, 
and $u_{A}(\wh{\emptyset}) = 0$.  In $\FdHilb$, where $V_o(A)$ is the space of hermitian operators on $A$, the trace is a unit.

Since $u_A$ is strictly positive, $\Omega(A) := u^{-1}(1) \cap V(A)_{+}$ is a 
{\em base} for the cone $V(A)_+$, i.e., every $\alpha \in V(A)_+$ is uniquely a non-negative multiple 
of a point in $\Omega(A)$ (\cite{AT}, Theorem 1.47) 
If $V(A)$ is finite-dimensional, this guarantees that $u$ is an {\em order unit} for $V(A)^{\ast}$ 
(as this space is spanned by evaluation functionals associated with elements $a \in \C(A,I)$), and we can regard $(V(A),V(A)^{\ast},u_A)$ as 
a COM. If $V(A)$ is infinite-dimensional, the matter is more delicate.  
Condition (ii) in Definition 5.1 does guarantee that 
$u_A$ will be an order unit for $V_{o}(A)^{\#}$, {\em if} $u_A$ belongs to this space, i.e., is a 
linear combination of functionals corresponding to elements of $\C(A,I)$.  It's worth recording 
the following corollary to Theorem 4.1.

{ \begin{corollary} Let $u$ be a unit for $V_o$, such that $u_A \in V_{o}^{\#}(A)$ for all $A \in \C$. Then 
\begin{itemize} 
\item[(a)] $(V_o(A), V_{o}^{\#}(A), u_A)$ is a COM for every $A \in \C$, and 
\item[(b)] $(V_o(A \otimes B), V_{o}^{\#}(A \otimes B), u_{A \otimes B})$ is a composite of 
$(V_o(A), V_{o}^{\#}(A), u_A)$ and $(V_o(B), V_{o}^{\#}(B), u_B)$ for every $A, B \in \C$. 
\end{itemize} 
\end{corollary} 
}

In general, however, $u_A$ will {\em not} belong to $\E(A)$. Indeed, if $\C = \Hilb$, the category of complex 
Hilbert spaces and bounded linear mappings, then $\V(A)$ can be identified with  the space of {\em finite-rank} operators on $A$. We can also identify each $a \in \E(A)$ with a finite-rank operator, with $a(\alpha) = \Tr(a \alpha)$ for 
all $\alpha \in \V(A)$. The natural 
choice of unit here is the trace functional $\alpha \mapsto \Tr(\alpha)$, but this corresponds to 
the identity operator on $A$, which isn't finite-rank. 

In such a situation, one can enlarge both $V_o(A)$ and $V_{o}(A)^{\#}$ so as to obtain a COM $(V(A), V^{\#}(A), u_A)$.
In fact, there are several different ways in which to do this. 
   

Given a unit $u$ for $V_o$, define an {\em effect} to be an element $a \in V_{o}(A)^{\ast}$ with 
$0 \leq a \leq u_A$ (in the dual ordering). As discussed earlier, an effect represents a mathematically 
possible measurement-outcome, since, for any normalized state 
$0 \leq a(\alpha) \leq 1$, so that we can regar $a(\alpha)$ as a probability. 
We write $[0,u_A]$ for the set of effects of $A$. 
There is a natural linear mapping $\ev : V_o(A) \rightarrow {\Bbb R}^{[0,u_A]}$, given by evaluation (that is, 
$\ev(\alpha)(a) = a(\alpha)$ for $a \in [0,u_A]$ and $\alpha \in V_o(A)$). By condition (i) in the definition of a unit,  
this mapping is injective. Henceforward, we shall identify $V_o(A)$ with 
its image under $\ev$, that is, we now regard $V_o(A)$ as a subspace of ${\Bbb R}^{[0,u_A]}$.\footnote{More exactly, 
there is a canonical surjection, given by restriction, from $\R^{[0,u_A]}$ to $\R^{\C(A,I)}$; this takes 
the image of $V_o(A)$ in the former isomorphically onto its image in the latter.}

\begin{definition}{\em Let $u$ be a unit for $V_o$. For each $A \in \C$, let $\Omega_o(A) = u_{A}^{-1}(1) \cap V_o(A)_+$. 
Then 
\begin{mlist} 
\item[$\bullet$] $\Omega(A,u)$ denotes the closure of $\Omega_o(A,u)$ in the product topology on ${\Bbb R}^{[0,u_A]}$. 
Note that this is a {\em compact} convex set.  
\item[$\bullet$]  $V(A)$ denotes the span of $\Omega(A,u)$ in $\R^{[0,u_A]}$, ordered by the cone 
$V(A)_+$ consisting of nonnegative multiples of points of $\Omega(A,u)$. 
That is, $V(A)_+  :=  \{ t\alpha  ~|~ \alpha \in \Omega(A,u) \ \mbox{and} \ t \geq 0\}$. 
\end{mlist}
}
\end{definition}

Since $\Omega(A,u)$ is compact, it follows that $V(A)$ is complete in the {\em base norm}, i.e, the 
Minkowski functional of the convex hull of $\Omega(A,u) \cup -\Omega(A,u)$. For details, see \cite{AlfsenShultzBook}.

\begin{lemma} The positive cone $\Vin(A)_{+}$ of $\Vin(A)$ is the pointwise closure, in ${\Bbb R}^{[0,u_A]}$, of $V_o(A)_{+}$. \end{lemma}

\noindent {\em Proof:} Let $K$ denote the closure of $V_o(A,u_A)$ in ${\Bbb R}^{[0,u_{A}]}$. Clearly, $V(A)_+ \subseteq K$. To establish the reverse inclusion, 
choose a net $t_i \rho_i \in V_{o}(A)_+$ with $\rho_i \in \Omega_o(A)$, and suppose $t_i \rho_i \rightarrow \rho \in {\Bbb R}^{[0,u_A]}$ in the product topology, i.e., pointwise: $\rho(a) = \lim_i t_i \rho_i(a) \geq 0$ for all $a \in [0,u_A]$. 
If $\rho(a) > 0$ for some $a$, then eventually we must have $t_i \rho_i(a) > 0$, whence, $\rho_i(a) > 0$. Now,  
$t_i = u_A(t_i \rho_i) \rightarrow u_A(\rho) = \rho(u_A)$. If $u_A(\rho) = 0$, then $t_i \rightarrow 0$. I claim this implies $\rho = 0$. If not, then for some $a \in [0,u_A]$ and $0 < \epsilon < \rho(a)$, 
we eventually have $t_i\rho_i(a) > \epsilon$, whence, as $a \leq u_A$, 
$t_i = (t_i \rho_i)(u_A) \geq t_i \rho_i(a) > \epsilon$, contradicting the fact that $t_i \rightarrow 0$.  Now 
suppose $\rho \not = 0$, so that $\rho(u_A) > 0$. Then we may write $\rho$ as $t \nu$ where $\nu = \rho/u_A(\rho)$ and $t = u_A(\rho)$; we then have $t_i = t_i \rho_i (u_A) \rightarrow t = \rho(u_A)$ and $t_i \rho_i(a) \rightarrow t \rho(a)$, whence, $\rho_i(a) \rightarrow \rho(a)$, for every $a \in [0,u_A]$. Thus, $\nu \in \Omega(A,u)$ and 
$\rho \in \Vin(A)_{+}$. $\Box$

\begin{definition} Let $\Eu(A,u)$ denote the span in $\Vin(A)^{\ast}$ of the evaluation functionals associated with points $a \in [0,u_A]$, ordered pointwise on $\Omega(A,u)$ (i.e., regarded as an ordered subspace of $\V(A)^{\ast}$ in the dual ordering). 
\end{definition}  

\noindent	{\em Notation:} From this point forward, let's assume a fixed unit $u$ is given, and, accordingly, abbreviate $\Vin(A,u)$ as $\Vin(A)$ and $\Eu(A,u)$ as $\Eu(A)$. Also, wherever it seems safe to do so, let's write $a(\alpha)$ for $\alpha(a)$, conflating $a \in [0,u_A]$ with the corresponding evaluation functional in $\Vin(A)^{\ast}$.

\begin{lemma} $(\Vin(A), \Eu(A), u_A)$ is a convex operational model. \end{lemma} 

\noindent{\em Proof:} $\Eu(A)$ is a separating space of functionals on $\Vin(A)$, and, by construction, 
$u_A$ is an order unit in $\Eu(A)$. $\Box$ \\

In any ordered abelian group, an interval $[0,u]$ is an {\em effect algebra} \cite{Foulis and Bennett} under the partial operation  $a \oplus b = a + b$ (defined for $a,b \in [0,u]$ provided that $a + b$ is again in $[0,u]$). 
In particular, for every $A \in \C$, $[0,u_A] \subseteq V_o(A)^{\ast}$ is an effect algebra. 

\begin{definition} A (finitely additive) {\em measure} on an effect algebra $L$ is a mapping $\mu : L \rightarrow {\Bbb R}_{+}$ such that, 
for all $a, b \in L$, $a \perp b \ \Rightarrow \ \mu(a \oplus b) = \mu(a) + \mu(b)$ (where $a \perp b$ means that $a \oplus b$ is defined). 
A {\em signed measure} on $L$ is a difference of measures.\end{definition} 

The set $M(L)$ of all signed measures on $L$ is a complete base-normed space; the dual 
order-unit is given by $u(\mu) = \mu(1_L)$ where $1_L$ is the unit element of $L$. It is easy to see that every element of $\Vin(A,u)$ is a signedd measure on the effect algebra $[0,u_A]$. Hence, $\Vin(A)$ is a closed subspace of $M(A) := M([0,u_A])$. 



\begin{example}{\em Let's  consider what these constructions yield where $\C = \Hilb$, with 
unit given by the trace.   
Let $A \in \Hilb$. As discussed above, $\V(A)$ is the space of finite-rank self-adjoint operators on $A$. 
Any positive linear functional $a \in V_o(A)^{\ast}$ with $a(\rho) \leq \tr(\rho)$ for all $\rho \in V_o(A)$ is bounded with respect to the trace norm on $V_o(A)$. Since $V_o(A)$ is trace-norm dense in the space 
--- let us denote it by $V_1(A)$ --- 
of self-adjoint trace-class operators, $a$ extends uniquely to a bounded linear functional on $V_1(A)$,   
whence, by a familiar duality, corresponds to a bounded self-adjoint operator $\hat{a}$ on $A$, given by $\tr(\rho \hat{a}) = a(\rho)$. 
Thus, we can identify $[0,u_A]$ with the standard interval of effects for $A$, and 
$\Eu(A)$ with the space of bounded self-addjoint operators on $A$. Finally, the extension of Gleason's Theorem to finitely additive measures on $[0,u_A]$ \cite{Christensen} allows us to identify $M(A)$ with the space spanned by the the set of finitely additive states on the factor $\B(A)$. Since the state space of $\B(A)$ is the weak-$\ast$ closed convex hull of the set of vector states, i.e., of the set of pure states in $V_o(A)$ (see, e.g., \cite{KR} Corollary 4.3.10), Lemma 5.4 
tells us that, in the cases arising in quantum theory, $\Vin(A) = M(A)$.} \end{example} 

This discussion suggests a third way in which we can complete $\V(A)$: 

\begin{definition} Let $\Omega_1(A,u)$ denote the closure of $\Omega_o(A,u)$ in the base norm on $\Vin(A,u)$, 
and let $V_1(A,u)$ be the 
closed subspace of $\Vin(A)$ spanned by $\Omega_1$ (and ordered by $V_1(A)_{+} = V_{1}(A) \cap \Vin(A)_{+}$). \end{definition}
 
If $\C = \Hilb$, with $u$ the trace, then $V_1(A)$ is self-adjoint trace class on $A$, so that our notation 
is consistent. 

It is easy to see that $M(A)$ is pointwise-closed in ${\Bbb R}^{[0,u_A]}$, whence, we have natural embeddings 
\[V_o(A) \leq V_1(A) \leq \Vin(A) \leq M(A) \leq {\Bbb R}^{[0,u_A]}.\]

Since the choice of $u_A$ is not canonical, we can't expect any of the constructions $V, M$ or $V_1$ to be functorial on $\C$. However, we can single out the sub-category of $\C$ having the same objects, but only those morphisms 
the images of which under $V_o$ are ``sub-normalizing" with respect to the unit: 

\begin{definition} 
$\C_{u}(A,B)$ consists of those morphisms $\phi \in \C(A,B)$ such that $(V_o\phi)^{\ast} ([0,u_B]) \subseteq [0,u_A]$ --- equivalently, such that $V_o(\phi)^{\ast}(u_B) = u_B \circ V_o(\phi) \leq u_A$. \end{definition}

It is straightforward that the composite (in $\C$) of morphisms $\phi \in \C_{u}(A,B)$ and $\psi \in \C_{u}(B,C)$ yields a morphism in $\C_{u}(A,C)$, so we have here a sub-category, $\C_u$, of $\C$. Moreover, since $\phi \in \C_{u}(A,B)$ implies that $V_o(\phi)^{\ast}([0,u_B]) \subseteq [0,u_A]$, we have a functor 
$M : \C_{u} \rightarrow \OrdLin$ given by 
\[M(\phi)(\mu)  = \mu \circ V_o(\phi)^{\ast}\]
where $\mu \in M(A)$ and $\phi \in \C(A,B)$.  In fact,  $\Vin$ and $V_1$ are also functorial 
with respect to $\C_u$:

\begin{lemma} $A \mapsto \Vin(A)$ and $A \mapsto V_1(A)$ are the object parts of functors $\Vin, V_1 : \C_u \rightarrow \OrdLin$, 
\end{lemma}

\noindent{\em Proof sketch:} Let $\phi \in \C_u(A,B)$. Then if $b \in [0,u_B]$, we have 
\[V_o(\phi)^{\ast}(b)(\rho) = b(V_o(\phi)(\rho)) \leq u_B(V_o(\phi)(\rho)) \leq u_A(\rho) \] 
for all $\rho \in V_o(A)_{+}$. 
Thus, $V_o(\phi)^{\ast}(b) \in [0,u_A]$.  We now have a continuous mapping ${\Bbb R}^{[0,u_A]} \rightarrow {\Bbb R}^{[0,u_B]}$, namely $V_o(\phi)^{\ast \ast} : \rho \mapsto \rho \circ V(\phi)^{\ast}$.  
It is straightforward that this mapping takes $\Omega_o(A,u_A)$ into $V_o(B,u_B)_{+}$; as it preserves effect-wise limits, it takes the effect-wise closure, $\Omega(A,u_A)$, of $\Omega_{o}(A,u_A)$ into the effect-wise closure 
of $V_o(B,u_B)_+$, which, by Lemma 5.4, is $\Vin(B,u)_+$. This gives us the desired positive linear mapping 
$\Vin(\phi) : \Vin(A) \rightarrow \Vin(B)$. 

To define $V_1(\phi)$, observe that since $u_B(\Vin(\phi)(\alpha)) \leq u_{A}(\alpha)$ for all $\alpha \in \Omega(A,u)$, we have 
$\|V(\phi)\| \leq 1$, where $\| \cdot \|$ denotes the operator norm, computed relative to the base norms 
on $V(A)$ and $V(B)$. In particular, $\Vin(\phi)$ is bounded, hence, continuous, with respect to these 
norms. Since $V(\phi)$ takes $V_o(A)_{+}$ into $V_o(B)_{+}$, it 
takes the span of the base-norm closure of the former cone to that of the latter, i.e, maps $V_1(A)$ into $V_1(B)$, giving us the desired positive linear mapping $V_1(\phi)$. 
$\Box$ 


\begin{corollary} If $V_o$ is monoidal, so are $\Vin$ and $V_1$. In particular, if $V_o(\C)$ is locally tomographic or $\C$ is compact closed, then both $\Vin$ and $V_1$ are 
monoidal functors. \end{corollary} 

\noindent{\em Proof:} Let $\phi \in \C(A,C)$ and $\psi \in \C(B,D)$. We wish to show that 
$\Vin(\phi \otimes \psi)$ depends only on $\Vin(\phi)$ and $\Vin(\psi)$. Let $\rho \in V_o(A \otimes B)$, 
$f \in [0,u_{CD}]$. We have 
$\Vin(\phi \otimes \psi)(\rho)(f) = \rho(f \circ V_o(\phi \otimes \psi))$ . 
But, if $V_o$ is monoidal, 
$V_o(\phi \otimes \psi)$ depends only on $V_o(\phi)$ and $V_o(\psi)$ --- whence, 
only on $\Vin(\phi)$ and $\Vin(\psi)$ (since $\Vin(\phi) = \Vin(\phi')$ implies $V_o(\phi) = V_o(\phi')$). 
The case of $V_1$ follows. $\Box$ \\

It remains to ask whether the pairs $(\Vin(A \otimes B), \Eu(A \otimes B), u_{A \otimes B})$ and 
$(V_1(A \otimes B), V_{1}^{\#}(A \otimes B), u_{A \otimes B})$ are respectable non-signaling composites, in the sense of definition 2.4.  I believe this to be the case, but do not have a proof. For the time being,  I leave this as a 
conjecture.

\section{Further Questions} 

This has been only a preliminary excursion into what looks like a rather large territory, raising many more questions than have been settled. Besides the conjecture mentioned above, a very partial list of unfinished business includes: (1) How are representations $V_{o}$ arising from 
various different monoid homomorphisms $\C(I,I) \rightarrow \R_+$ related to one another? 
(2) If $\C$ is dagger compact, and $V : \C \rightarrow \Rvec$ is a monoidal representation, will the category 
$V(\C)$ be weakly self-dual in the sense of \cite{BDW}? (3) How do the constructions sketched above (notably, $V_o$) interact with Selinger's CPM construction \cite{Selinger}?  (4) Can one characterize abstractly those symmetric monoidal categories $\C$ 
for which (there exists a monoid homomorphism $p : \C(I,I) \rightarrow \R_+$ such that) $\C$ is isomorphic, or equivalent, to $V_o(\C)$? (5) What is the connection between the constructions described here and the approach to constructing 
operational models based on Chu spaces, explored in \cite{Abramsky, Abramsky-Heunen}? 

Ultimately, the convex operational models considered here, are less basic, and less flexible, than 
probabilistic models associated with test spaces \cite{BW, Wilce}. It would be extremely 
interesting to know how to define something like a test space associated with each object in a symmetric monoidal 
category, in purely category-theoretic terms.  One candidate is the set of special commuative $\dagger$-Frobenius 
algebras associated with the given object. The question then arises: what is the 
image of such an algebra under a representation, e.g., $V_o$? 

 
\providecommand{\urlalt}[2]{\href{#1}{#2}}
\providecommand{\doi}[1]{doi:\urlalt{http://dx.doi.org/#1}{#1}}

\end{document}

%% file: discardingtree.tex
\newif\ifvflip\pgfkeys{/tikz/vflip/.is if=vflip}
\newif\ifhflip\pgfkeys{/tikz/hflip/.is if=hflip}
\newif\ifhvflip\pgfkeys{/tikz/hvflip/.is if=hvflip}
\newlength\morphismheight
\newlength\wedgewidth
\tikzset{width/.initial=1mm}
\makeatletter
\tikzstyle{morphism}=[font=\small,morphismshape]
\pgfdeclareshape{morphismshape}
{
    \savedanchor\centerpoint
    {
        \pgf@x=0pt
        \pgf@y=0pt
    }
    \anchor{center}{\centerpoint}
    \anchorborder{\centerpoint}
    \saveddimen\overallwidth
    {
        \pgfkeysgetvalue{/tikz/width}{\minwidth}
        \pgf@x=\wd\pgfnodeparttextbox
        \ifdim\pgf@x<\minwidth
            \pgf@x=\minwidth
        \fi
    }
    \savedanchor{\upperrightcorner}
    {
        \pgf@y=.5\ht\pgfnodeparttextbox
        \advance\pgf@y by -.5\dp\pgfnodeparttextbox
        \pgf@x=.5\wd\pgfnodeparttextbox
    }
    \anchor{north}
    {
        \pgf@x=0pt
        \pgf@y=0.5\morphismheight
    }
    \anchor{north east}
    {
        \pgf@x=\overallwidth
        \multiply \pgf@x by 2
        \divide \pgf@x by 5
        \pgf@y=0.5\morphismheight
    }
    \anchor{east}
    {
        \pgf@x=\overallwidth
        \divide \pgf@x by 2
        \advance \pgf@x by 5pt
        \pgf@y=0pt
    }
    \anchor{west}
    {
        \pgf@x=-\overallwidth
        \divide \pgf@x by 2
        \advance \pgf@x by -5pt
        \pgf@y=0pt
    }
    \anchor{north west}
    {
        \pgf@x=-\overallwidth
        \multiply \pgf@x by 2
        \divide \pgf@x by 5
        \pgf@y=0.5\morphismheight
    }
    \anchor{south east}
    {
        \pgf@x=\overallwidth
        \multiply \pgf@x by 2
        \divide \pgf@x by 5
        \pgf@y=-0.5\morphismheight
    }
    \anchor{south west}
    {
        \pgf@x=-\overallwidth
        \multiply \pgf@x by 2
        \divide \pgf@x by 5
        \pgf@y=-0.5\morphismheight
    }
    \anchor{south}
    {
        \pgf@x=0pt
        \pgf@y=-0.5\morphismheight
    }
    \anchor{text}
    {
        \upperrightcorner
        \pgf@x=-\pgf@x
        \pgf@y=-\pgf@y
    }
    \backgroundpath
    {
    \begin{scope}
        \pgfkeysgetvalue{/tikz/fill}{\morphismfill}
        \pgfsetstrokecolor{black}
        \pgfsetlinewidth{.7pt}
        \begin{scope}
        \pgfsetstrokecolor{black}
        \pgfsetfillcolor{white}
        \pgfsetlinewidth{.7pt}
                \ifhflip
                    \pgftransformyscale{-1}
                \fi
                \ifvflip
                    \pgftransformxscale{-1}
                \fi
                \ifhvflip
                    \pgftransformxscale{-1}
                    \pgftransformyscale{-1}
                \fi
                \pgfpathmoveto{\pgfpoint
                    {-0.5*\overallwidth-5pt}
                    {0.5*\morphismheight}}
                \pgfpathlineto{\pgfpoint
                    {0.5*\overallwidth+5pt}
                    {0.5*\morphismheight}}
                \pgfpathlineto{\pgfpoint
                    {0.5*\overallwidth + \wedgewidth}
                    {-0.5*\morphismheight}}
                \pgfpathlineto{\pgfpoint
                    {-0.5*\overallwidth-5pt}
                    {-0.5*\morphismheight}}
                \pgfpathclose
                \pgfusepath{fill,stroke}
        \end{scope}
    \end{scope}
    }
}
\makeatother

\newenvironment{pic}[1][]
{\begin{aligned}\begin{tikzpicture}[font=\tiny,#1]}
{\end{tikzpicture}\end{aligned}}

\pgfdeclarelayer{foreground}
\pgfdeclarelayer{background}
\pgfdeclarelayer{morphismlayer}
\pgfsetlayers{background,main,morphismlayer,foreground}

\newlength\minimummorphismwidth
\newlength\stateheight
\newlength\minimumstatewidth
\newlength\connectheight
\tikzset{colour/.initial=white}

\tikzstyle{pure}=[line width=.7pt]

\makeatletter

\pgfdeclareshape{ground}
{
    \savedanchor\centerpoint
    {
        \pgf@x=0pt
        \pgf@y=0pt
    }
    \anchor{center}{\centerpoint}
    \anchorborder{\centerpoint}

    \anchor{north}
    {
        \pgf@x=0pt
        \pgf@y=0.16\stateheight
    }
    \anchor{south}
    {
        \pgf@x=0pt
        \pgf@y=0pt
    }
    \saveddimen\overallwidth
    {
        \pgfkeysgetvalue{/pgf/minimum width}{\minwidth}
        \pgf@x=\minimumstatewidth
        \ifdim\pgf@x<\minwidth
            \pgf@x=\minwidth
        \fi
    }
    \backgroundpath
    {
        \begin{pgfonlayer}{foreground}
        \pgfsetstrokecolor{black}
        \pgfsetlinewidth{1.25pt}
        \ifhflip
            \pgftransformyscale{-1}
        \fi
        \pgftransformscale{0.5}
        \pgfpathmoveto{\pgfpoint{-0.5*\overallwidth}{0pt}}
        \pgfpathlineto{\pgfpoint{0.5*\overallwidth}{0pt}}
        \pgfpathmoveto{\pgfpoint{-0.33*\overallwidth}{0.33*\stateheight}}
        \pgfpathlineto{\pgfpoint{0.33*\overallwidth}{0.33*\stateheight}}
        \pgfpathmoveto{\pgfpoint{-0.16*\overallwidth}{0.66*\stateheight}}
        \pgfpathlineto{\pgfpoint{0.16*\overallwidth}{0.66*\stateheight}}
        \pgfpathmoveto{\pgfpoint{-0.02*\overallwidth}{\stateheight}}
        \pgfpathlineto{\pgfpoint{0.02*\overallwidth}{\stateheight}}
        \pgfusepath{stroke}
        \end{pgfonlayer}
    }
}

\newcommand{\tinyground}[1][ground]{
\smash{\raisebox{-2pt}{\hspace{-3pt}\ensuremath{\begin{pic}[scale=0.4]
    \node[#1, scale=0.6] (1) at (0,0.4) {};
    \draw [pure] (1.south) to +(0,-.3);
\end{pic}
}\hspace{-1pt}}}}

\newcommand{\discard}[1]{\ensuremath{\tinyground_{#1}}}